\documentclass[10pt,conference,letterpaper]{IEEEtran}
\IEEEoverridecommandlockouts
\usepackage{cite}
\usepackage{amsmath,amssymb,amsfonts}
\usepackage{graphicx}
\usepackage{algorithmicx}
\usepackage{booktabs}
\usepackage[noend]{algpseudocode}
\usepackage{subfloat}
\usepackage{textcomp}
\usepackage{xcolor}
\usepackage{enumitem}
\usepackage{array}
\usepackage[font=footnotesize]{caption}
\usepackage{footnote}
\usepackage{url}
\usepackage{balance}
\usepackage{multirow}
\usepackage{hyperref}
\usepackage{gensymb}
\usepackage[caption=false,font=footnotesize]{subfig}
\def\BibTeX{{\rm B\kern-.05em{\sc i\kern-.025em b}\kern-.08em
    T\kern-.1667em\lower.7ex\hbox{E}\kern-.125emX}}
\newcolumntype{b}{X}
\newcolumntype{s}{>{\hsize=.5\hsize}X}

\makeatletter
\newcommand\fs@spaceruled{\def\@fs@cfont{\bfseries}\let\@fs@capt\floatc@ruled
  \def\@fs@pre{\vspace*{0.2cm}\hrule height.8pt depth0pt \kern2pt}%
  \def\@fs@post{\kern2pt\hrule\relax}%
  \def\@fs@mid{\kern2pt\hrule\kern2pt}%
  \let\@fs@iftopcapt\iftrue}
\makeatother

\begin{document}

\title{Starlink on the Road: A First Look at\\ Mobile Starlink Performance in Central Europe}

\author{ 
Dominic Laniewski\textsuperscript{1}, Eric Lanfer\textsuperscript{1}, Simon Beginn\textsuperscript{1}, Jan Dunker\textsuperscript{1}, Michael Dückers\textsuperscript{2}, Nils Aschenbruck\textsuperscript{1}\\

\textsuperscript{1}Osnabrück University - Institute of Computer Science, Osnabrück, Germany\\
\textsuperscript{2}SWO Netz GmbH - Osnabrück, Germany\\
\{laniewski, lanfer, sbeginn, jdunker, aschenbruck\}@uos.de, michael.dueckers@swo-netz.de\\
}

\maketitle

\begin{abstract}
Low Earth Orbit Satellite Networks such as Starlink promise to provide world-wide Internet access. 
While traditionally designed for stationary use, a new dish, released in April 2023 in Europe, provides mobile Internet access including in-motion usage, e.g., while mounted on a car.
In this paper, we design and build a mobile measurement setup. Our goal is to fully autonomously conduct continuous Starlink measurements while the car is in motion. We share our practical experiences, including challenges regarding the permanent power supply. We measure the Starlink performance over the span of two months from mid-January to mid-March 2024 when the car is in motion. The measurements consist of all relevant network parameters, such as the download and upload throughput, the RTT, and packet loss, as well as detailed power consumption data. We analyze our dataset to assess Starlink's mobile performance in Central Europe, Germany, and compare it to stationary measurements in proximity.
We find that the mobile performance is significantly worse than stationary performance. The power consumption of the new dish is higher, but seems to be more correlated to the heating function of the dish than to the speed of the vehicle.
\begingroup%
\renewcommand\thefootnote{}
\footnote{Dominic Laniewski and Eric Lanfer are co-first authors}%
\addtocounter{footnote}{-1}
\endgroup%
\end{abstract}

\begin{IEEEkeywords}
Starlink, Satellite Communication, Mobility, LEO, Measurement, Dataset, Starlink Mobility, Starlink Dataset, Starlink Measurement
\end{IEEEkeywords}

\section{Introduction} \label{sec-intro}
In recent years, Low Earth Orbit (LEO) satellite networks have gained an increasing popularity because they promise to provide consumer-level Internet access around the world. Today, SpaceX provides the largest LEO satellite network -- Starlink. Since its release for public access in 2020, users can gain access via stationary Starlink dishes, typically mounted on rooftops of buildings. SpaceX has released a new dish, called \emph{Flat High Performance} dish, in December 2022 in the US and in April 2023 in Europe. We refer to this dish as \emph{Starlink FHP} dish in the following. It is designed to be used in-motion, e.g., when mounted on a vehicle's roof (such as boat roof and car roof).

The understanding of Starlink's functioning and achievable performance is still an ongoing research challenge. Most studies aim to get a first understanding of the achievable performance in stationary use-cases through measurements \cite{kassem2022browser,ma2023network,michel2022first,datasetpaper} or simulations \cite{lai2020starperf,kassing2020exploring,cao2023satcp}. First recent studies also assess Starlink's mobile performance while mounted on a car in the US \cite{hu2023leo} and in the Arctic \cite{beckman2024mobile}. While providing detailed performance assessments, they either lack a clear description of the physical measurement setup, or the setup is designed for non-autonomous short-term deployments.

In this paper, we mount a Starlink FHP dish on a car, conduct continuous measurements of Starlink's performance while the car is driving in Central Europe, and analyze the resulting dataset.
Our contributions are as follows:
\begin{itemize}
    \item We design and build a mobile Starlink measurement setup. The goal is to fully autonomously conduct continuous Starlink measurements while the car is in motion.
    \item We share practical experiences and open challenges with our setup, especially regarding permanent power supply of the dish.
    \item We use this system to measure Starlink's performance over the span of two months from mid-January until mid-March 2024, when the car is in motion in a city in Central Europe, Germany. The measurements include all relevant network parameters such as download and upload throughput, RTT, and packet loss, as well as detailed power consumption data. 
    \item We analyze our dataset to assess Starlink's mobile performance and compare it to stationary measurements in spatial proximity. We find that the mobile performance is significantly worse than stationary performance. The power consumption of the FHP dish is higher, but seems to be more correlated to the heating function of the dish than to the speed of the vehicle.
    \item We make an anonymized version of our dataset publicly available \cite{dataRed}.
\end{itemize} 

The remainder of this paper is structured as follows. First, we discuss related work in Sec.~\ref{sec-related-work} and provide background information about Starlink in Sec.~\ref{sec-background}. Afterward, we present our mobile measurement setup, including our physical mounting solution, as well as the tooling used, in Sec.~\ref{sec-mobile-setup}. In Sec.~\ref{sec-performance}, we take a first look at mobile Starlink performance by analyzing our collected data and comparing it to stationary measurements. Afterward, we discuss open challenges in Sec.~\ref{sec-discussion}. We conclude our paper in Sec.~\ref{sec-conclusion}.

\section{Related Work} \label{sec-related-work}

Lately, many studies aimed to assess Starlinks performance in stationary use-cases through simulations or real-world measurements.
\emph{Hypatia} \cite{kassing2020exploring} and \emph{StarPerf} \cite{lai2020starperf} are frameworks that simulate the network characteristics based on satellite behavior.
Michel et al. \cite{michel2022first} take a first look at Starlink performance by measuring throughput via TCP and QUIC, latency, and packet loss. 
Kassem et al. \cite{kassem2022browser} use a browser plugin to measure web-performance over a Starlink link.
Zhao et al. \cite{zhao2023realtime} measure Starlink performance for real-time multimedia services such as video-on-demand, live video streaming, and video conferencing.
Ma et al. \cite{ma2023network} measure throughput via TCP and UDP, latency, packet loss, and routing information from urban and outback vantage points.
Tiwar et al. \cite{tiwari2023t3p} conduct latency measurements from the north (UK) and south (Spain) of Europe.
Raman et al. \cite{raman2023dissecting} compare Starlink performance to Geostationary Equatorial Orbit (GEO) and Medium Earth Orbit (MEO) satellite networks.
Pan et al. \cite{pan2023measuring} conduct throughput, latency, and traceroute measurements and analyze the Starlink point of presence (POP) structure.
Garcia et al. \cite{garcia2023multi} analyze frequency scheduling and beam switching via throughput measurements.
Izhikevich et al. \cite{izhikevich2023democratizing} conduct world-wide latency measurements by probing publicly exposed services that are connected via Starlink.
Mohan et al. \cite{mohan2023multifaceted} assess Starlinks global throughput performance by analyzing the M-Lab speedtest dataset.
Laniewski et al. \cite{datasetpaper} provide a large-scale, publicly available dataset of Starlink measurements including throughput, latency, packet loss, traceroute, and weather data collected over a span of six months from Enschede (The Netherlands) and Osnabrück (Germany).

Recent studies also measure Starlink's mobile performance.
López et al. \cite{lopez2023connecting} mount a Starlink Gen-1 dish on a car and measure latency at a velocity of 15~km/h on a 250~km test drive through a rural area in northern Denmark.
Similarly, Ma et al. \cite{ma2023network} also mount a Starlink Gen-1 dish on a van. They drove for 30 minutes at a velocity around 40 - 70~km/h in south-western Canada and measured download and upload throughput, as well as latency. 
Compared to stationary measurements, they found largely similar throughput performances, but significant latency spikes and fluctuations.
Hu et al. \cite{hu2023leo} mount both, a Starlink Gen-2 and Starlink FHP on a car. They measure TCP and UDP upload and download throughput, latency and packet loss on a trip over 3,800~km across five states in the US with velocities up to 100~km/h. They compare the performance of Starlink's roaming and mobility plans and compare their results to cellular measurements.
Beckman et al. \cite{beckman2024mobile} mount a Starlink mobile dish on a van and conduct TCP download throughput measurements and latency measurements on a two-days long 970~km test drive across the arctic region in northern Sweden at velocities of 80-100~km/h. Their analysis focuses on the impact of Starlink's 15-second reconfiguration interval on the download throughput.

While these existing studies provide extensive measurements of Starlink's mobility performance in different regions of the world -- primarily the US and Arctic, they lack a clear description of their measurement setup, especially regarding the continuous power supply of the dish. To the best of our knowledge, we are the first to share practical experiences and challenges of creating a fully autonomous Starlink mobility measurement system. Using this system we conduct mobility measurements in Central Europe, Germany, consisting of all relevant network parameters such as download and upload throughput, RTT, and packet loss, as well as detailed power consumption data. Our results validate existing measurements and gain new insights into Starlink's power consumption. 
\section{Starlink Overview} \label{sec-background}

Starlink has been opened for public access in 2020.
Users access the network via a specialized satellite antenna called the "dishy". It is traditionally mounted at a stationary place such as a rooftop of a building. The recently released \emph{Flat High Performance} (FHP) dish is designed to be mounted on vehicles such as boats and cars to allow in-motion mobile usage.
The dish connects to a satellite, which, on the other hand, is also connected to a Ground Station (GS). This link (dish - satellite - GS) is called the one-hop bent-pipe. It can be extended to a multi-hop bent-pipe if no GS is in the coverage of the satellite. In this case, newer satellites, beginning with version 1.5, can form inter-satellite links (ISL) using laser communication.
The GS is connected to a terrestrial point of presence (POP) structure.

Starlink offers different dishes for different use-cases. The \emph{Standard} dish is primarily designed for stationary use but can also be used for in-motion scenarios with speeds up to 16~km/h \cite{starlinkplanoverview}. It has an average power usage of 50-75~W and a field of view (FOV) of 100° \cite{starlinkdishspecs}. The achievable performance differs for different regions in the world, but SpaceX advertises downlink throughput of up to 220~Mbit/s, upload throughput of up to 25~Mbit/s, and latencies in the range of 25-60~ms \cite{starlinkplanspecs}.
The Starlink mobile dish is designed for in-motion use. SpaceX advertises up to 220~Mbit/s, up to 25~Mbit/s, and $< 99$~ms in download and upload throughput, and latency, respectively \cite{starlinkplanspecs}. Furthermore, it has an average power usage of 110-150~W and a FOV of 140° \cite{starlink-spec-flathighperf}.

SpaceX offers standard and priority traffic plans. Priority traffic is prioritized in times of network congestion, leading to better performance.

\section{The Mobile Measurement Setup} \label{sec-mobile-setup}

In this section, we describe our mobile measurement setup. First, we define our unique measurement scenario and constraints. Afterwards, we describe our physical measurement setup. Then, we describe the measurement process, which is the software-side of our setup. It includes details such as the tools and configurations we used to measure Starlink performance.

\begin{figure}
    \centering
    \includegraphics[width=\linewidth]{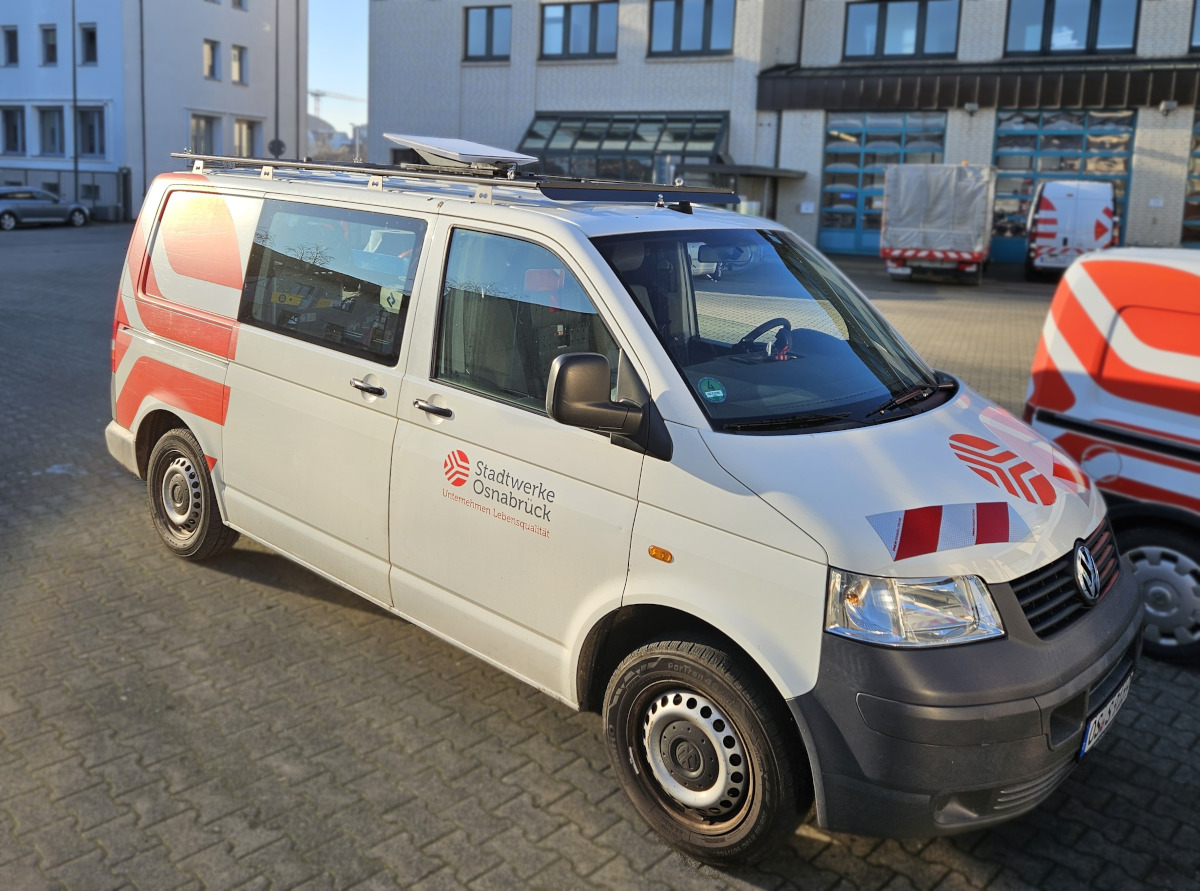}
    \caption{The van with the Starlink FHP dish}
    \label{fig:assembly-whole-van}
\end{figure}

\subsection{The Measurement Scenario and Constraints}
We teamed-up with an energy infrastructure provider (SWO Netz GmbH) and mounted the Starlink dish on the roof of one of their service vans. These vans drive to clients distributed all over the city, sometimes including the rural districts. This scenario is fundamentally different to related work, who drove large distances in single journeys \cite{hu2023leo,beckman2024mobile} with the only purpose to measure Starlink performance. In contrast, our van typically drives short distances of a few kilometers at a time and at irregular intervals. It is likely that it doesn't drive at all on some days (e.g., the weekends), and a few different routes on other days, with potentially arbitrarily long intervals between the routes where the car is turned off. Furthermore, we faced the unique challenge that the driver of the car is no expert in the field, does not know details about our measurement setup, and thus, can only provide limited maintenance.

These unique circumstances led to our goal to build an autonomous measurement system that conducts continuous Starlink measurements while the car is in-motion. The setup should be suitable for a long-term deployment over multiple weeks with as little required manual maintenance as possible, imposing unique challenges for safety and power supply.
Generally, the setup needs to automatically boot up and start the measurements when the car is turned on, and automatically stop the measurements and turn off when the car is turned off. The van with the mounted Starlink dish is depicted in Fig.~\ref{fig:assembly-whole-van}.

\subsection{The Physical Measurement Setup}
\begin{figure}
    \centering
    \includegraphics[width=\linewidth]{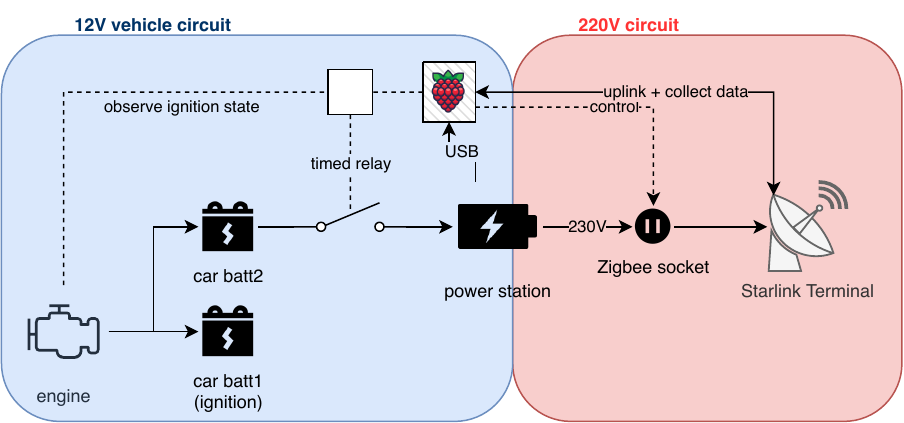}
    \caption{Overview of our physical measurement setup.}
    \label{fig:setup_car}
\end{figure}

\begin{figure}
    \centering
    \includegraphics[width=\linewidth]{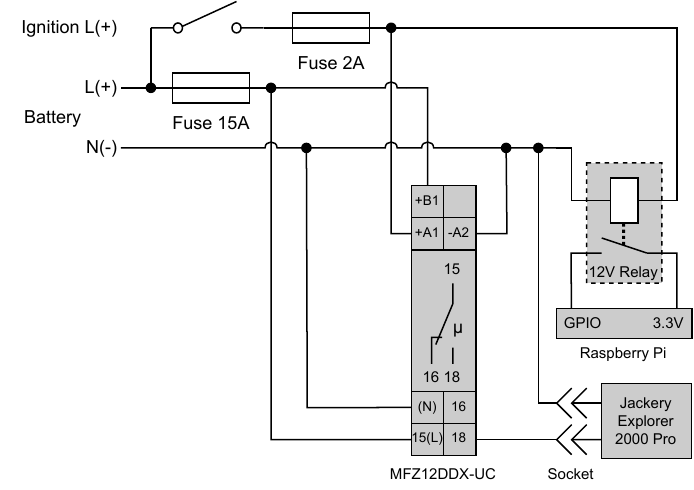}
    \caption{Wiring diagram}
    \label{fig:wiring-diagram}
\end{figure}

\begin{figure}
    \centering
    \includegraphics[width=\linewidth]{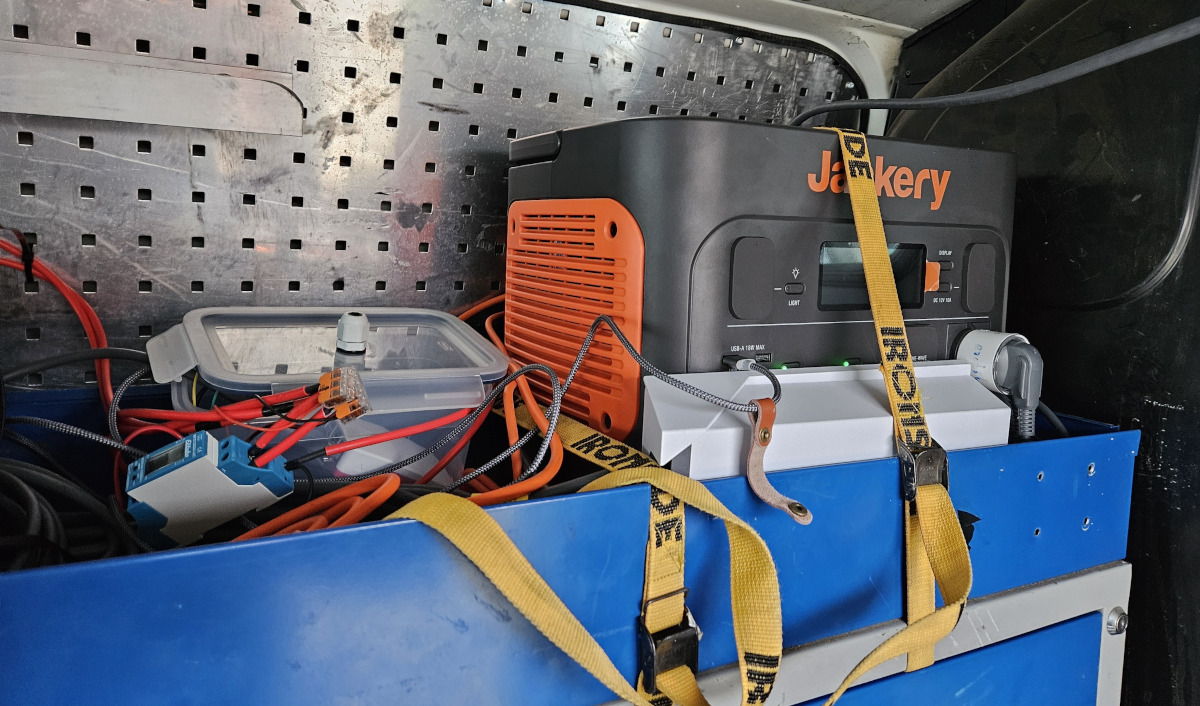}
    \caption{Raspberry Pi, relays, connections to car battery and ignition, Jackery Explorer 2000 Pro with smart Zigbee socket and Starlink FHP dish power supply.}
    \label{fig:assembly-setup}
\end{figure}
An overview of our physical measurement setup is depicted in Fig.~\ref{fig:setup_car}.
The service van provides an internal 12~V DC power system and has two car batteries installed to guarantee reliability.
We use a Jackery Explorer 2000 Pro power station with 2~kWh capacity as central power supply for our setup. 
It has a built-in inverter and supports both, 230~V AC wall charging and 12~V DC charging from the car battery.
It is connected via a timed relay (Eltako MFZ12DDX-UC) to the second car battery. 
This relay is also connected to the ignition of the car to detect when the car engine is turned on and off. The power station automatically starts to be loaded from the car battery when the engine is turned on. The relay is programmed to a fixed time interval (in our case 15 minutes), in which the power station is still loaded after the engine is turned off.

The Starlink dish is connected via a Zigbee smart socket (NOUS A1Z) to one of the 230V AC outputs of the power station.
We use the smart socket to monitor the power consumption of the dish, as well as to turn the dish on and off.

Furthermore, we deploy a Raspberry Pi 4 as the main controller of our measurements. It not only runs our measurement software, but also turns the Starlink dish on and off. It has an attached USB Zigbee dongle (Sonoff 3.0 USB Zigbee Dongle Plus) to communicate with the Zigbee smart socket. Moreover, it observes the ignition state using a second relay (Omron G2R-1-E DC12) and turns the smart socket off if the engine is off, and on if the engine is on.
Fig.~\ref{fig:wiring-diagram} provides a detailed overview of the wiring of the relays. Especially the Omron relay is notable. It signals the engine state to the Raspberry Pi by completing the circuit from a $3.3 V$ pin to a GPIO pin. The internal pull down resistor of the GPIO pin is enabled, so the voltage will drop off when the relay opens. We opted to use Home Assistant\footnote{\url{https://www.home-assistant.io}} to automate the controlling of the smart socket.
Lastly, appropriately rated fuses were added to both the ignition signal line and the car battery line to protect the electronics in case of a fault. 
Fig.~\ref{fig:assembly-setup} shows a picture of the final setup inside the service van.

\begin{figure}
    \centering
    \includegraphics[width=\linewidth]{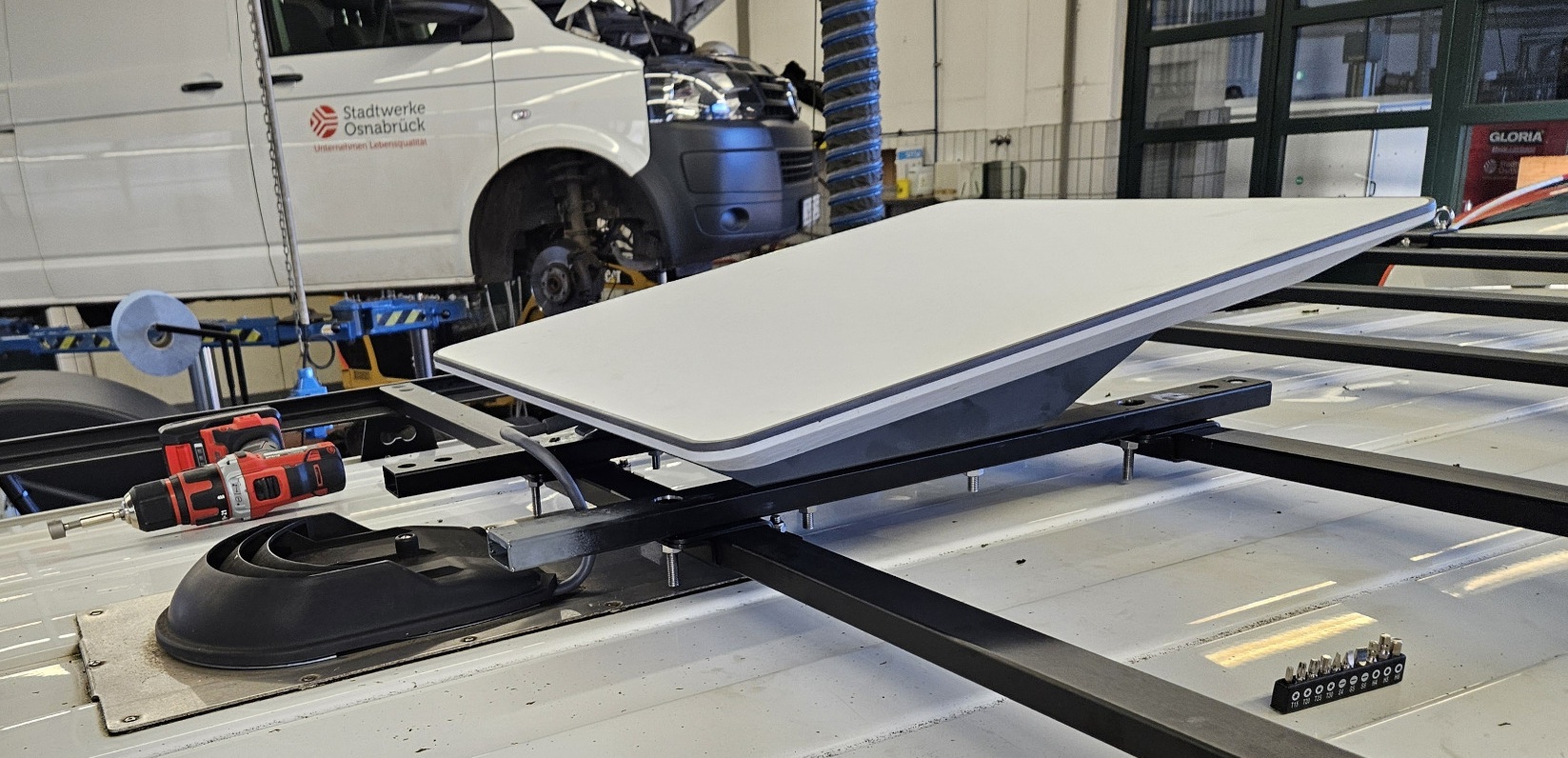}
    \caption{The Starlink dish mounted on the roof rack.}
    \label{fig:assembly-roofmount}
\end{figure}

The Starlink FHP dish comes by default with the Wedge Mount Kit. Most notably, the kit includes a baseplate that can be bolted to a flat surface and holds up the dish at a slight angle.
The service van comes with a waterproof vent designed to pass through cables from the rooftop to the inside. We mounted the Starlink dish with the baseplate on top of a roof rack for safe operation also in case of a crash at high velocity. Our final mount, including the cable attached to the dish, can be seen in Fig.~\ref{fig:assembly-roofmount}.

\begin{figure}
    \centering
    \includegraphics[width=\linewidth]{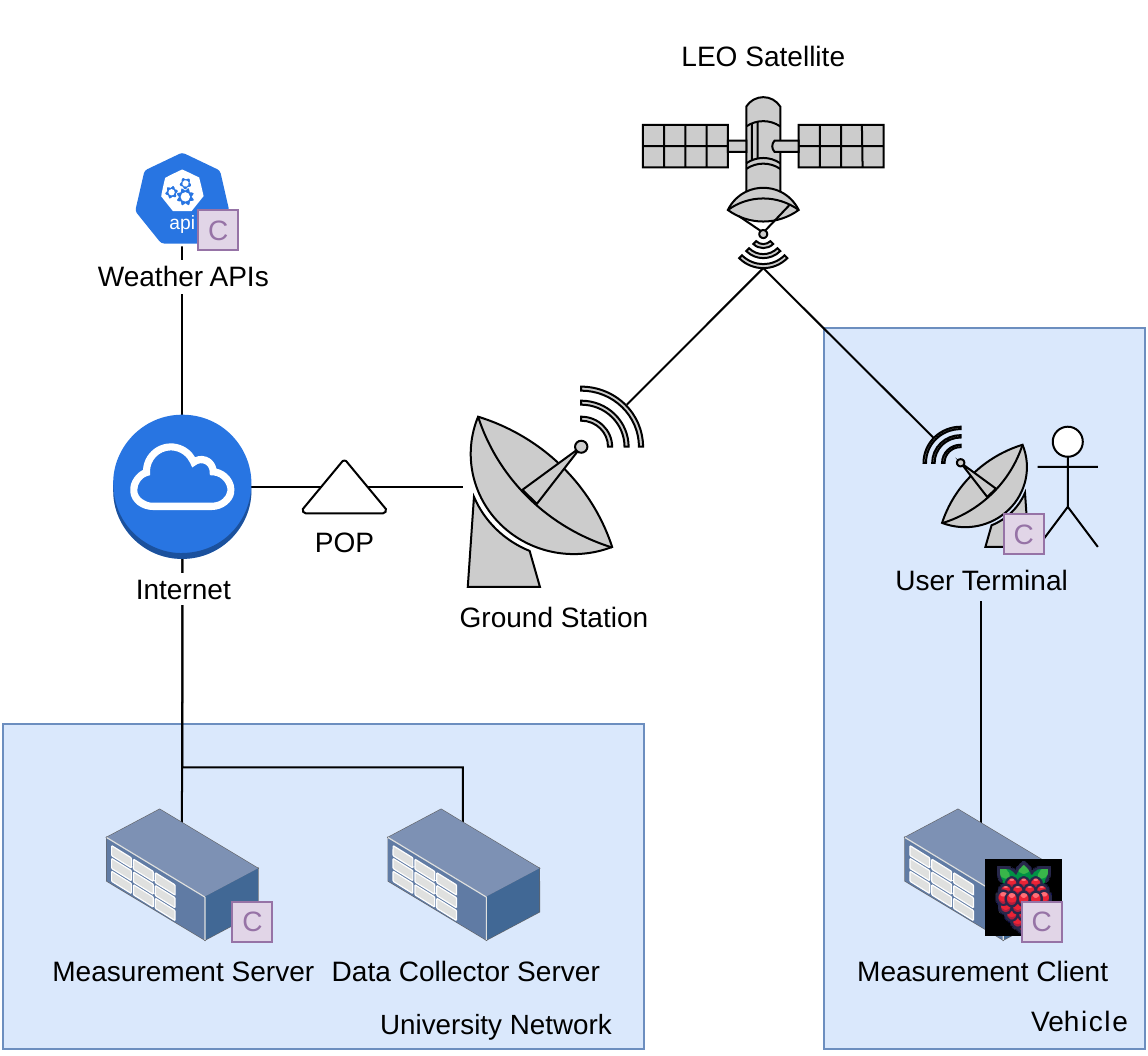}
    \caption{Diagram of the measurement setup. The purple $C$ indicates data endpoints, that
are collected by the data collection server. }
    \label{fig:leo-diagram}
\end{figure}
\begin{figure}
    \centering
    \includegraphics[width=\linewidth]{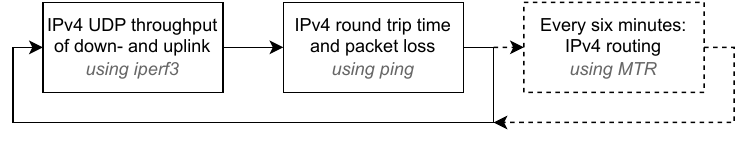}
    \caption{The network measurement process. Based on \cite{datasetpaper}.}
    \label{fig:cycle}
\end{figure}

\subsection{The Measurement Process} 
Fig.~\ref{fig:leo-diagram} depicts our measurement setup. It is a modified version of the setup presented in \cite{datasetpaper}. 
The RaspberryPi 4 deployed in the van coordinates the measurements. It is connected via gigabit Ethernet to the Starlink router, which is set into bypass mode. The measurements are conducted against a measurement server located in the university network, which has a shared 5~Gbit/s WAN connection to the Internet. The results are transmitted via the same Starlink connection to the data collector server. We acknowledge that this might affect the performance measurements, but since the amount of data required for this task is minimal, this effect is negligible. Based on publicly available information \cite{starlink-sx} we reason that the traffic likely reaches the Ground Station in Aerzen, Germany, and is routed over Starlinks' POP in Frankfurt, Germany.
The following key performance indicators are collected: download and upload throughput, RTT, packet loss, and power consumption of the dish. Furthermore, we collect weather data from the nearest-by Deutscher Wetterdienst (DWD) weather station (ID 00342).

We measure Starlink performance using the iterative process visualized in Fig.~\ref{fig:cycle}. We measure UDP throughputs and RTTs with packet loss at intervals of less one minute. Approximately every six minutes we additionally collect traceroutes.
Since we are interested in raw performance, we measure the downlink and uplink throughputs with UDP. We start two parallel \emph{iperf3} (version 3.9) instances to avoid possible CPU limitations of iperfs' bidirectional mode and measure for approximately 15 seconds with target datarates of 500~Mbit/s in downlink and 100~Mbit/s in uplink.
Afterwards, we measure the RTT and packet loss using the \emph{ping} tool by sending 250 packets at intervals of 0.1~s. This measurement cycle repeats continuously. Approximately every six minutes we additionally collect traceroutes using \emph{Matt's traceroute} (MTR, version 0.94) \cite{mtr} with 15 report cycles.

In addition to the network parameters, the smart Zigbee socket enables us to log the current power consumption of the Starlink dish. We record the state of the socket, the current, voltage and power in a 10 second interval.

Furthermore, we collect the diagnostic data provided by the Starlink dish. This includes the firmware version, obstruction, and longitude and latitude coordinates. They are reported at 30 second intervals.

As we use the Starlink connection also to transmit the results to the collector server, we store the results locally on the Pi in case no Starlink connection is available and retransmit them once a connection is available again.

\section{Analysis} \label{sec-performance}
In this section, we analyze our dataset. First, we provide a general overview of the dataset. Afterward, we analyze the spatial distribution of Starlink performance. Thereafter, we analyze the impact of speed on the throughput, RTT, as well as packet loss and compare our results to stationary measurements. At the end, we take a look at the power consumption of the Starlink dish. 

\begin{figure}
    \centering
    \subfloat[Samples per location]{
    \includegraphics[width=0.46\linewidth]{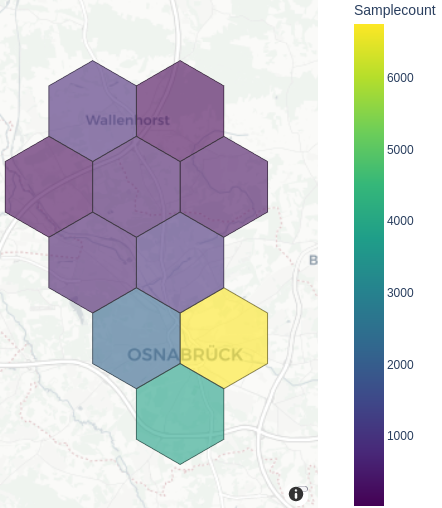}\label{fig:map_mobile_location}
    }
    \subfloat[Download Throughput]{
    \includegraphics[width=0.49\linewidth]{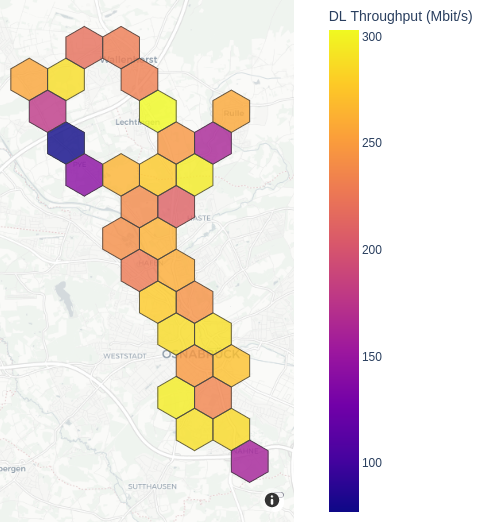}\label{fig:map_mobile_speed}
    }
    \caption{Map visualization of our measurements locations}
\end{figure}

\subsection{The Dataset}
Our dataset, collected from January, 11th until the March, 7th, 2024, consists of $15,269$ measurement samples. The van was in motion in $4,196$ (27\%) of the measurements. As our subscription includes 50\,GB priority traffic, we have $5,180$ samples measured with priority conditions. We compared the priority and non-priority measurements in terms of throughput. We get download throughput medians of 284.68~Mbit/s and 303.85~Mbit/s for non-priority and priority, respectively. Overall, non priority is approximately 6\% slower than priority traffic. For upload throughput, non-priority traffic is approximately 8\% slower, with medians of 16.45~Mbit/s and 17.91\,Mbit/s, respectively. To have a larger baseline of samples, and as the differences are small, we will not distinguish priority and non-priority measurements for the core part of the analysis, assuming that the higher throughput is constant for the priority traffic. However, at different vantage points in the analysis, we will provide details on the priority aspects, when necessary. 

The local distribution of the data is visualized in Fig.~\ref{fig:map_mobile_location}. The map shows that most of the samples were collected in the inner city of Osnabrück, the base of the electric grid provider is located in the yellow hexagon with the most samples. The three hexagons with the highest distribution of samples are all located in a densely populated area, the hexagons in the north display the Osnabrück county region, which is more rural.

\subsection{Spatial Distribution of the Network Performance}
Assuming that the signal in densely populated sectors is more likely obstructed by buildings, we analyze the download throughput based on the visualization in Fig.~\ref{fig:map_mobile_speed}. In the north-west, we observe a cluster of three hexagons with lower throughput, the rest of the map is more or less equally distributed with higher throughput values without preference of urban or less urban areas. With a closer look at the low throughput cluster, we argue that these measurements were close to the local mountain, with a height of 188~m, which probably obstructed the signal path, especially when the van was in downhill positions. However, our dataset does not provide data on the orientation and inclination of the van to further proof this assumption. Further, we analyzed the round-trip-time and the packet loss rates in terms of their local distribution. Close to the mentioned mountain, we observed high loss rates (about 45\%) and high RTTs with 349~ms. In terms of the RTT, we cannot observe any further abnormalities. Regarding the loss distribution, we observed loss rates of 1\% across the measurement region. Especially in the inner-city region, we can observe some sectors with loss rates up to 10\%. We argue that these higher loss rates were induced by obstacles hindering the direct line of sight to the satellites, as we took a closer look at the building structure in these sectors. This observation is inline with~\cite{hu2023leo}, where a worse throughput was measured in urban and suburban area types compared to rural areas.

\begin{figure}
    \centering
    \subfloat[Download]{
    \includegraphics[width=0.49\linewidth]{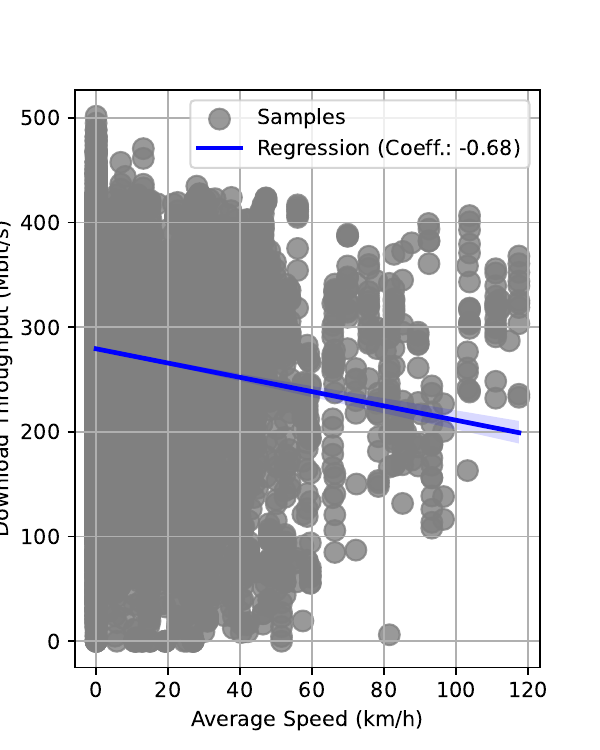}\label{fig:dl_scatter}
    }
    \subfloat[Upload]{
    \includegraphics[width=0.49\linewidth]{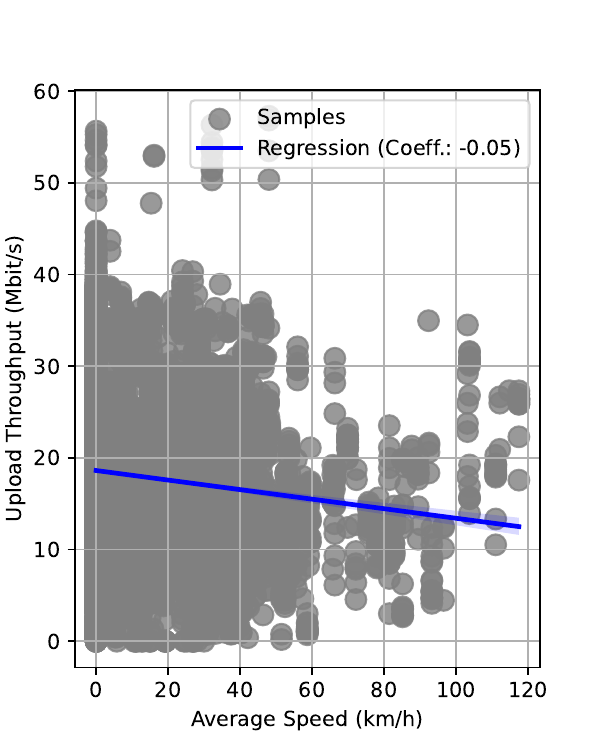}\label{fig:ul_scatter}
    }
    \caption{Throughput vs. average moving speed.}
\end{figure}

\subsection{The Impact of Speed}

Next, we discuss the impact of speed on Starlink performance. As an overview of our dataset, we present scatter plots of the download and upload throughput vs. speed in Fig.~\ref{fig:dl_scatter} and Fig.~\ref{fig:ul_scatter}. Since most of the dataset was recorded in a city, a big part of the samples is located at speed below $50$\,km/h, which is the inner-city speed limit in Germany. Inspecting the regression line for the download throughput, we observe a stronger correlation than for the upload throughput, suggesting a connection that higher speeds degrade the download throughput rate. However, as a limiting factor of this connection, we have fewer samples in the higher speed areas. Moreover, since the measurements were taken inside the city, it is likely that a degradation was caused by shadowing of buildings while being in movement. 

\begin{figure}
    \centering
    \includegraphics[width=1\linewidth]{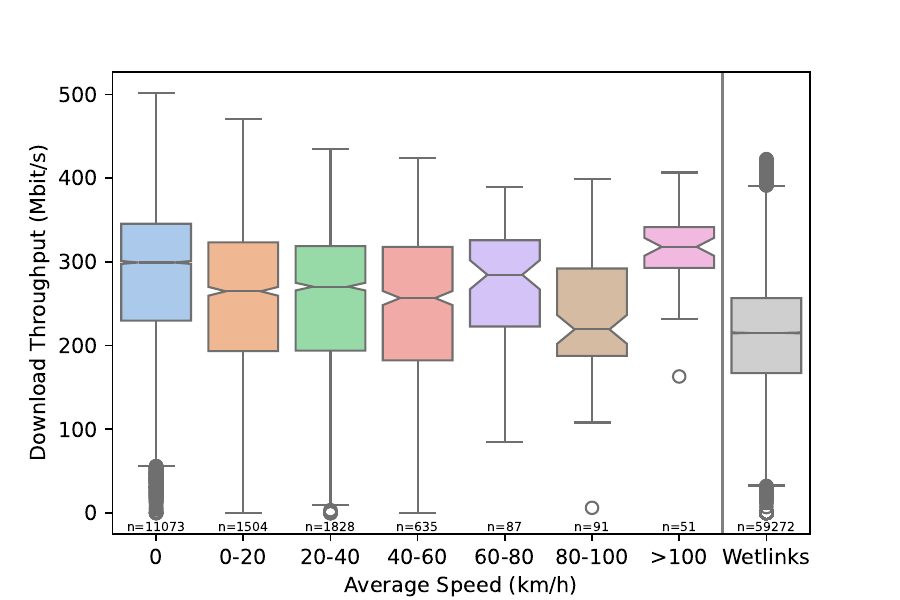}
    \caption{Download throughput compared between different speed groups and the stationary Wetlinks dataset (station Osnabrück) \cite{datasetpaper}}
    \label{fig:speed_buckets}
\end{figure}

\begin{figure}
    \centering
    \includegraphics[width=1\linewidth]{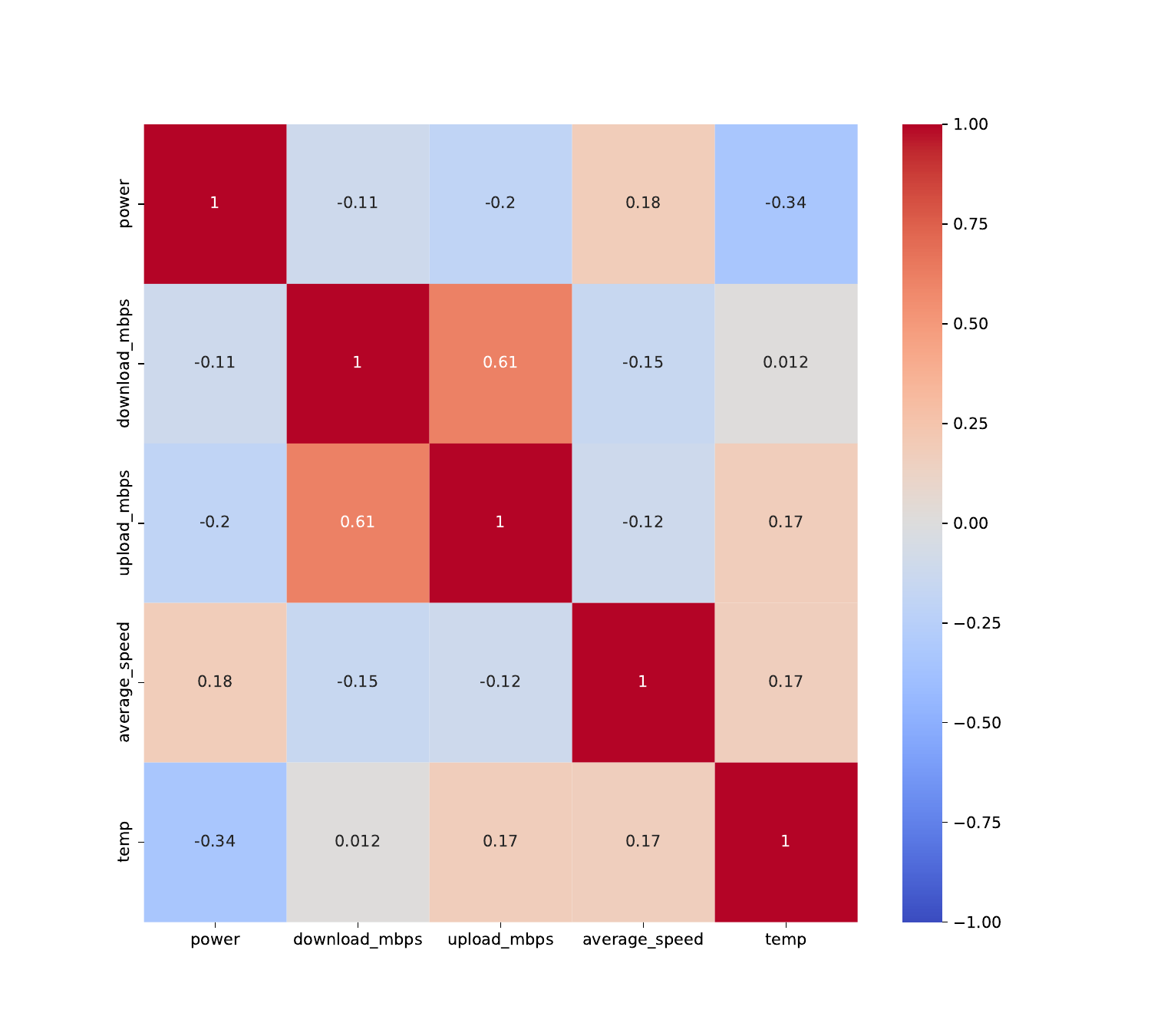}
    \caption{Correlation matrix}
    \label{fig:corr}
\end{figure}

To further investigate this issue, we grouped our samples into buckets by speed and present the data in boxplots in Fig.~\ref{fig:speed_buckets}. We observe a median download throughput of 299\,Mbit/s for a standing vehicle, 265\,Mbit/s at speeds up to 20\,km/h, 270\,Mbit/s at speeds up to 40\,km/h, and a fluctuating one for speeds higher than 40\,km/h. These fluctuations are caused by the decreasing number of samples for higher speeds. Our results indicate that the vehicle speed does not have a direct effect on the download throughput. However, the download throughput is approximately 10\% lower, if the car is generally in motion, compared to if it is standing still. 
This finding is also supported by a correlation analysis, as visualized in Fig.~\ref{fig:corr}. We only observe low correlation coefficients for the impact of speed on download ($-0.15$) and upload ($-0.12$) throughput. 
For speeds higher than 100\,km/h we observe a higher throughput. We note that the number of samples is quite low and see this as an area of future work. However, we believe that the higher throughput is likely correlated to a clear direct line of sight to the satellites. In Germany, speeds higher than 100\,km/h are typically measured on the Autobahn, where there are typically no buildings close-by that affect the throughput negatively.

Fig.~\ref{fig:speed_buckets} also includes the WetLinks data \cite{datasetpaper} from the stationary measurement station in Osnabrück for comparison. The observed download throughput at speeds of 0~km/h significantly outperforms the WetLinks data by approximately 70~Mbit/s in the median. Since our measurement setup is comparable to the one used in the WetLinks dataset, this difference is likely primarily caused by the different dish versions, indicating a significantly higher performance of the new FHP dish compared to the stationary Gen-2 dish.

Similarly to the download throughput, we observe a smaller median upload throughput in motion with 15.40\,Mbit/s, compared to 17.74\,Mbit/s for a standing vehicle. The previously named limitations, induced by measurements in the city, do also apply here. Concerning the priority non-priority issue, we computed median download throughput for all samples in motion. With priority subscription, a median throughput of 277.89\,Mbit/s was achieved (n=1097), without subscription it was at 262.77\,Mbit/s (n=3099).

In addition to the throughput, we analyzed the loss and RTT values at certain speed levels. 
We computed a correlation coefficient for the worst ping samples of a measurement run vs. average speed, indicating a very fragile correlation of $0.13$. However, further analysis indicated that we do not have enough samples at higher speeds to validate this correlation properly. When analyzing the influence of speed on loss, the correlation coefficient is $0.05$, indicating almost no impact. 

In related work, Hu et al. \cite{hu2023leo} reported an average in-motion UDP download throughput of 128~Mbit/s, performing significantly worse than our results. We suspect that this is primarily caused by the different measurement locations. Their measurements were conducted in the US, where Starlink is known to achieve significantly worse performance compared to Central Europe \cite{starlinkperformanceoverview}. Hu et al. also report that the network performance is largely unaffected by movement speed. While our data shows similar behavior for in motion measurements, we observe a noticeable and statistically significant drop in download throughput between stationary and in motion measurements. This suggests that once in motion, the speed doesn't have a noticeable impact on network performance.

\subsection{Power consumption}
With the use of the Zigbee power socket, we were able to collect data on the power consumption. Overall, the dish had an average power consumption of 113\,W, a minimum of 48\,W, and a maximum of 191\,W. This average consumption is in line with the specification provided by the company itself \cite{starlink-spec-flathighperf}. However, compared to the 50-75\,W consumed by the standard actuated dish, this consumption amount of the FHP dish is quite high. We also included the power consumption in the correlation matrix in Fig.~\ref{fig:corr}. The matrix indicates an influence of temperature on the power consumption with a correlation coefficient of $-0.34$. We analyzed it further and found evidence that the consumption is higher at low-temperature levels, especially below 0\,C°. However, this observation does only hold true for a limited number of samples. We assume that this is caused by punctual use of the heating function provided by the dish. The second-highest correlation with $0.18$ is the impact of speed on power consumption. Analyzing all samples where the van was in motion, we observe a median power consumption of 137\,W. In static moments, we observe a median power consumption of 106\,W. This observation, does not necessarily provide a causality between movement and power consumption because the dish usually consumes more power, while starting up. The starting up will mostly happen after the engine was started, and the vehicle gets in motion. In the static measurements (e.g., at the end of a drive), the dish is likely already connected, and the measurements are running after the engine was turned off. 

\section{Practical Experiences and Open Challenges} \label{sec-discussion} 
During our two months measurement campaign, we gained practical experiences and identified open challenges for our measurement setup.

First, the Jackery Explorer 2000 Pro power station was not perfectly suited for our use-case. Our measurement campaign was conducted during winter time, with temperatures below $0^\circ C$ on some days. The power station has a minimum charge temperature of $0^\circ C$\footnote{\url{https://r.jackery.net/productGuide/Jackery\%20Explorer\%202000\%20Pro\%20Tragbare\%20Powerstation\%20Benutzerhandbuch.pdf}}, meaning that it could not be charged on these days. 
Furthermore, it turns off the power to outlets after a certain time period, if the power draw is low. It can only be turned on again by manually pushing buttons on the station. We found that the built-in low power mode, that is advertised to prevent this behavior, does not function correctly. To keep this behavior from occurring, it is necessary to artificially increase the power consumption for each socket periodically. Therefore, we ran a 10 seconds stress test on the Raspberry Pi and booted the Starlink dish for 1.5 minutes every three hours to reset the turn-off timer of the power station. This results in unwanted power drain in periods where the car is turned off, e.g., the weekends or during nighttime.

The main open challenge is the continuous power supply of the FHP dish. We measured an average consumption of 113~W, and peak consumption of 191~W. The internal 12~V vehicle circuit can charge the power station with approximately 90~W leading to a consistent power drain of the power station. This problem is further amplified by the periodical stress tests and dish boot-ups to prevent a complete shutdown of the system.

\section{Conclusion} \label{sec-conclusion}
In this paper, we built a mobile Starlink measurement setup with the goal to fully autonomously conduct continuous Starlink measurements while a vehicle is in-motion. We mounted the new Starlink FHP dish on the roof of a service van of an energy infrastructure provider and conducted Starlink measurements over the span of two months from mid-January until mid-March 2024 in a city in Central Europe, Germany. These measurements contain all relevant network parameters, such as download and upload throughput, RTT, packet loss, as well as power consumption data of the dish. We made an anonymized version of the resulting dataset publicly available and analyzed it to assess Starlink mobile performance.

Our analysis results suggest that the download and upload throughput rates drop by approximately 10\% when the vehicle is in motion. Once in motion, the speed of the car has no further impact on the throughput. Moreover, we observed higher loss rates in urban areas, likely caused by obstruction through buildings. Additionally, we observed a higher power consumption, while the vehicle is in motion and at cold temperature conditions. 

Finally, our measurement campaign has identified the continuous power supply of the Starlink dish as an open challenge for future work. We measured an average power consumption of 113~W with peaks up to 190~W while a car can provide a maximum of approximately 90~W. This means that our measurement setup currently cannot be operated fully autonomously, but occasional recharging of the power station is required.
\section{Acknowledgment} \label{sec-ack}
We thank the SWO Netz GmbH for letting us build our measurement setup into their van, allowing us to gather the data, and being very helpful with troubleshooting issues.
We like to thank Justus Bachmann, Alexander Böckenholt, Bennet Janzen, and Malte Wehmeier for their help implementing the measurement tools and collecting the data.
This work has been partially supported by the German Federal Ministry for Digital and Transport as part of the “Innovative Network Technologies” funding program (FKZ: 19OI23008C). 
\section{Ethical Statement}
The measurements were conducted carefully in line with the fair-use policy of the network provider. Moreover, to the best of our knowledge, our measurements did not interfere with the service for other users. Our data contains sensitive user data in the form of the latitude and longitude coordinates of the car, making it possible to track the driver as well as the customers. While the driver agreed that this data is used for our analysis, we did not track the exact routes driven and we base our analysis on time and/or location independent analyses such as heatmaps or location-independent speeds. Moreover, the public version of our dataset preserves anonymity in the form that we removed the GPS coordinates.

\balance
\bibliographystyle{IEEEtranS}
\bibliography{IEEEabrv, bibliography}

\begin{thebibliography}{10}
\providecommand{\url}[1]{#1}
\csname url@samestyle\endcsname
\providecommand{\newblock}{\relax}
\providecommand{\bibinfo}[2]{#2}
\providecommand{\BIBentrySTDinterwordspacing}{\spaceskip=0pt\relax}
\providecommand{\BIBentryALTinterwordstretchfactor}{4}
\providecommand{\BIBentryALTinterwordspacing}{\spaceskip=\fontdimen2\font plus
\BIBentryALTinterwordstretchfactor\fontdimen3\font minus
  \fontdimen4\font\relax}
\providecommand{\BIBforeignlanguage}[2]{{%
\expandafter\ifx\csname l@#1\endcsname\relax
\typeout{** WARNING: IEEEtranS.bst: No hyphenation pattern has been}%
\typeout{** loaded for the language `#1'. Using the pattern for}%
\typeout{** the default language instead.}%
\else
\language=\csname l@#1\endcsname
\fi
#2}}
\providecommand{\BIBdecl}{\relax}
\BIBdecl

\bibitem{beckman2024mobile}
C.~Beckman, J.~Garcia, H.~Mikkelsen, and P.~Persson, ``{Starlink and Cellular
  Connectivity under Mobility: Drive Testing Across the Arctic Circle},'' in
  \emph{Proceedings of the Wireless Telecommunications Symposium (WTS)}, 2024.

\bibitem{cao2023satcp}
X.~Cao and X.~Zhang, ``{SaTCP: Link-Layer Informed TCP Adaptation for Highly
  Dynamic LEO Satellite Networks},'' in \emph{Proceedings of the IEEE
  International Conference on Computer Communications (INFOCOM)}, 2023, pp.
  1--10.

\bibitem{garcia2023multi}
J.~Garcia, S.~Sundberg, G.~Caso, and A.~Brunstrom, ``{Multi-Timescale
  Evaluation of Starlink Throughput},'' in \emph{Proceedings of the 1st ACM
  Workshop on LEO Networking and Communication}, 2023, pp. 31--36.

\bibitem{hu2023leo}
B.~Hu, X.~Zhang, Q.~Zhang, N.~Varyani, Z.~M. Mao, F.~Qian, and Z.-L. Zhang,
  ``{LEO Satellite vs. Cellular Networks: Exploring the Potential for
  Synergistic Integration},'' in \emph{Proceedings of the 19th International
  Conference on emerging Networking EXperiments and Technologies (CoNEXT)},
  2023, pp. 45--51.

\bibitem{izhikevich2023democratizing}
L.~Izhikevich, M.~Tran, K.~Izhikevich, G.~Akiwate, and Z.~Durumeric,
  ``{Democratizing LEO Satellite Network Measurement},'' \emph{Proceedings of
  the ACM on Measurement and Analysis of Computing Systems}, vol.~8, no.~1,
  2024.

\bibitem{kassem2022browser}
M.~M. Kassem, A.~Raman, D.~Perino, and N.~Sastry, ``{A Browser-side View of
  Starlink Connectivity},'' in \emph{Proceedings of the 22nd ACM Internet
  Measurement Conference (IMC '22)}, 2022, p. 151–158.

\bibitem{kassing2020exploring}
S.~Kassing, D.~Bhattacherjee, A.~B. \'{A}guas, J.~E. Saethre, and A.~Singla,
  ``{Exploring the “Internet from space” with Hypatia},'' in
  \emph{Proceedings of the 20th ACM Internet Measurement Conference (IMC '20)},
  2020, p. 214–229.

\bibitem{lai2020starperf}
Z.~Lai, H.~Li, and J.~Li, ``{StarPerf: Characterizing Network Performance for
  Emerging Mega-Constellations},'' in \emph{Proceedings of the 28th IEEE
  International Conference on Network Protocols (ICNP)}, 2020, pp. 1--11.

\bibitem{dataRed}
D.~Laniewski, E.~Lanfer, S.~Beginn, J.~Dunker, M.~Dückers, and N.~Aschenbruck,
  ``{Starlink on the Road Dataset},''
  \url{https://github.com/sys-uos/Starlink-on-the-Road}, 2024.

\bibitem{datasetpaper}
D.~Laniewski, E.~Lanfer, B.~Meijerink, R.~van Rijswijk-Deij, and
  N.~Aschenbruck, ``{WetLinks: a Large-Scale Longitudinal Starlink Dataset with
  Contiguous Weather Data},'' \emph{arXiv preprint arXiv:2402.16448}, 2024.

\bibitem{lopez2023connecting}
M.~L{\'o}pez, S.~B. Damsgaard, I.~Rodr{\'\i}guez, and P.~Mogensen,
  ``{Connecting Rural Areas: An Empirical Assessment of 5G Terrestrial-LEO
  Satellite Multi-Connectivity},'' in \emph{Proceedings of the 97th IEEE
  Vehicular Technology Conference (VTC2023-Spring)}, 2023, pp. 1--5.

\bibitem{ma2023network}
S.~Ma, Y.~C. Chou, H.~Zhao, L.~Chen, X.~Ma, and J.~Liu, ``{Network
  Characteristics of LEO Satellite Constellations: A Starlink-Based Measurement
  from End Users},'' in \emph{Proceedings of the IEEE International Conference
  on Computer Communications (INFOCOM)}, 2023, pp. 1--10.

\bibitem{michel2022first}
F.~Michel, M.~Trevisan, D.~Giordano, and O.~Bonaventure, ``{A First Look at
  Starlink Performance},'' in \emph{Proceedings of the 22nd ACM Internet
  Measurement Conference (IMC '22)}, 2022, p. 130–136.

\bibitem{mohan2023multifaceted}
N.~Mohan, A.~Ferguson, H.~Cech, P.~R. Renatin, R.~Bose, M.~Marina, and J.~Ott,
  ``{A Multifaceted Look at Starlink Performance},'' in \emph{Proceedings of
  the ACM Web Conference (WWW '24)}, 2024.

\bibitem{pan2023measuring}
J.~Pan, J.~Zhao, and L.~Cai, ``{Measuring a Low-Earth-Orbit Satellite
  Network},'' in \emph{Proceedings of the 34th IEEE Annual International
  Symposium on Personal, Indoor and Mobile Radio Communications (PIMRC)}, 2023,
  pp. 1--6.

\bibitem{starlink-sx}
M.~Puchol, ``Starlink tracker,'' \url{https://starlink.sx/}, 2024.

\bibitem{raman2023dissecting}
A.~Raman, M.~Varvello, H.~Chang, N.~Sastry, and Y.~Zaki, ``{Dissecting the
  Performance of Satellite Network Operators},'' \emph{Proceedings of the ACM
  on Networking}, vol.~1, no. CoNEXT3, 2023.

\bibitem{starlinkperformanceoverview}
SpaceX, ``{Starlink Availability Map},''
  \url{https://www.starlink.com/map?source=roam&view=download}, 2024.

\bibitem{starlink-spec-flathighperf}
------, ``{Starlink Flat High Performance Kit Specifications},''
  \url{https://api.starlink.com/public-files/Starlink%20Product%20Specifications_FlatHighPerformance.pdf},
  2024.

\bibitem{starlinkplanoverview}
------, ``{Starlink Service Plan Overview},''
  \url{https://www.starlink.com/service-plans}, 2024.

\bibitem{starlinkplanspecs}
------, ``{Starlink Service Plan Specifications},''
  \url{https://www.starlink.com/legal/documents/DOC-1400-28829-70}, 2024.

\bibitem{starlinkdishspecs}
------, ``{Starlink Standard Kit Specifications},''
  \url{https://api.starlink.com/public-files/Starlink%20Product%20Specifications_Standard.pdf},
  2024.

\bibitem{tiwari2023t3p}
S.~Tiwari, S.~Bhushan, A.~Taneja, M.~Kassem, C.~Luo, C.~Zhou, Z.~He, A.~Raman,
  N.~Sastry, L.~Qiu, and D.~Bhattacherjee, ``T3p: Demystifying low-earth orbit
  satellite broadband,'' 2023.

\bibitem{mtr}
R.~Wolff, ``{Matt's Traceroute (MTR)},'' \url{https://www.bitwizard.nl/mtr/},
  2024.

\bibitem{zhao2023realtime}
H.~Zhao, H.~Fang, F.~Wang, and J.~Liu, ``{Realtime Multimedia Services over
  Starlink: A Reality Check},'' in \emph{Proceedings of the 33rd Workshop on
  Network and Operating System Support for Digital Audio and Video (NOSSDAV)},
  2023, p. 43–49.

\end{thebibliography}

\end{document}